\documentclass[reprint,aps,pre]{revtex4-1}
\usepackage{graphicx}

\newcommand{\vc}{\mathbf}

\begin{document}
\author{Rasmus A. X. Persson}
\affiliation{Department of Chemistry \& Molecular Biology, University of
Gothenburg, SE-412 96 Gothenburg, Sweden}
\email{rasmusp@chem.gu.se}
\title{Sigma method for the microcanonical entropy or density of states}

\begin{abstract}
We introduce a simple improvement on the method to calculate equilibrium
entropy differences between classical energy levels proposed by Davis (S.
Davis, Phys. Rev. E, 050101, 2011). We demonstrate that the modification is
superior to the original whenever the energy levels are sufficiently closely
spaced or whenever the microcanonical averaging needed in the method is carried
out by importance sampling Monte Carlo. We also point out the necessary
adjustments if Davis's method (improved or not) is to be used with molecular
dynamics simulations.
\end{abstract}



\maketitle

Consider a system with configurational coordinates $\{\vc r_i\}$
and potential energy function $U(\{\vc r_i\})$. The Hamiltonian of the system
is of the standard classical form, that is, separable in its coordinates and
conjugate momenta, $\{\vc p_i\}$, so that it may be written thus
\begin{equation}
H(\{\vc r_i\}, \{\vc p_i\}) = U(\{\vc r_i\}) + K(\{\vc p_i\})
\end{equation}
where $K(\{\vc p_i\})$ is the kinetic energy of the system. In this Brief
Report, we consider the calculation of the energy dependence of the
microcanonical Boltzmann-Planck equilibrium entropy,
\begin{equation}
\label{eq:planck}
S(E) = k \ln \omega(E)
\end{equation}
where $k$ is Boltzmann's constant and 
\begin{equation}
\label{eq:dos}
\omega(E) = C \int \{ d \vc r_i \} \{ d \vc p_i \} \delta(E - H(\{ \vc r_i \},
\{\vc p_i\}))
\end{equation}
is the phase density, also known as the density of states, where $C$ is a
constant that assures $\omega(E)$ is dimensionless and $\delta$ is Dirac's
$\delta$ function. Algorithms to evaluate $\omega(E)$ (and thus $S(E)$) abound
in the literature
\cite{ferrenberg88,labastie90,berg91,cheng92,poteau94,geyer95,calvo95,ming96,wang99,wang01a,adib05,heilmann05,skilling06,partay10,do11,persson12,davis13}.
They each have their advantages and drawbacks, and an exhaustive review of them
all is not possible in this Brief Report. Here, we instead focus in particular
on the recently proposed $\sigma$ method by Davis \cite{davis11}. We will
recapitulate its derivation and offer an improvement on the original method.
Our notation differs slightly from that of Davis.

For the moment, we consider a microcanonical ensemble whose only first integral
of motion is the total energy, $E$. We will briefly consider the case with more
first integrals of motion later. The Laplace principle of indifference assigns
equal \textit{a priori} probability to all phase space points on the energy
shell
$H(\{\vc r_i\}, \{\vc p_i\}) = E$. In other words, the ensemble probability
density is constant on this energy shell and zero everywhere else. We write
this probability density as
\begin{equation}
\label{eq:prob}
W_E(\{\vc r_i\}, \{\vc p_i\}) = \frac {C} {\omega(E)} \delta(E - H(\{\vc r_i\},
\{\vc p_i\}))
\end{equation}
If $K(\{\vc p_i\})$ is quadratic in each conjugate momentum coordinate and
shows no complicated interdependencies (in this equation $\{m_i\}$ are
generalized masses),
\begin{equation}
K(\{\vc p_i\}) = \sum_i \frac {\vc p_i^2} {2 m_i},
\end{equation}
the dependence on $\{\vc p_i\}$ can be integrated out, yielding
\cite{severin78, schranz91, ray91, davis11},
\begin{equation}
\label{eq:kdos}
\widehat W_E(\{\vc r_i\}) = C_E' \left (E - U(\{\vc r_i\}) \right )^{n / 2 - 1}
\Theta (E - U(\{\vc r_i\}))
\end{equation}
where
\begin{equation}
\label{eq:norm}
C_E' = \left ( \int \{ d \vc r_i \} (E - U(\{\vc r_i\}))^{n/2 -1} \Theta(E -
U(\{ \vc r_i \})) \right )^{-1}
\end{equation}
is a normalization constant that is inversely proportional to
$\omega(E)$, $\Theta$ is the Heaviside step function, and $n$ is the number of
configurational degrees of freedom of the system (which for an unconstrained
particle system is three times the number of particles in three dimensions). The
quantity $\widehat W_E(\{\vc r_i\})$ is directly proportional to the density of
kinetic energy states. The microcanonical average of a quantity $A(\{\vc
r_i\})$ that does not depend explicitly on the momenta can now be expressed as,
\begin{equation}
\label{eq:ave}
\langle A(\{\vc r_i\}) \rangle_E = \int \{d \vc r_i\} \widehat W_E(\{ \vc r_i
\}) A(\{ \vc r_i \}).
\end{equation}

The probability function in eq.~(\ref{eq:kdos}) can be used as the weighting
factor in a microcanonical Markov chain Monte Carlo simulation \cite{severin78,
schranz91, ray91} to calculate averages according to eq.~(\ref{eq:ave}). In a
molecular dynamics simulation, however, in which additional integrals of motion
appear, the probability function of eq.~(\ref{eq:kdos}) is not the proper one,
assumed ergodicity notwithstanding. In this case, the correct probability
function is given by \cite{ray99},
\begin{equation}
\widetilde W_E(\{\vc r_i\}) = \widetilde C_E \left (E - U(\{\vc r_i\}) - \frac
{\vc P^2} {2 M} \right )^{n/2 - 1}
\end{equation}
where $\vc P$ is the center-of-mass momentum, $M$ the total mass, $\widetilde
C_E$ a normalization constant and $n$ carries the same meaning as in
eq.~(\ref{eq:kdos}) but does not correspond to the same numerical value, there
being one degree of freedom less for each Cartesian component of the
center-of-mass momentum.

At this point, Davis \cite{davis11} introduces the quantity
\begin{equation}
\label{eq:davis}
\sigma_{E, E_\mathrm{r}}(\{\vc r_i\}) = \frac {\Theta(E_\mathrm{r} - U(\{\vc
r_i\}))} {(E - U(\{ \vc r_i \}))^{n / 2 - 1}}
\end{equation}
with the condition that $E_\mathrm{r} \leq E$, but $E_\mathrm{r}$ otherwise
arbitrary. When the microcanonical average
of eq.~(\ref{eq:davis}) is calculated using eq.~(\ref{eq:ave}) and
eq.~(\ref{eq:kdos}), keeping in mind that $C_E' \propto 1 / \omega(E)$, it is
seen that that the entropy difference according to eq.~(\ref{eq:planck})
between two energy levels $E'$ and $E''$ is given by, 
\begin{equation}
\label{eq:delta1}
\Delta_{E'}^{E''} S = k \ln \frac {\langle \sigma_{E', E_\mathrm{r}}(\{\vc
r_i\}) \rangle_{E'}} {\langle \sigma_{E'', E_\mathrm{r}}(\{\vc r_i\})
\rangle_{E''}},
\end{equation}
which is the rational for introducing the $\sigma$ function. Similar to Davis's procedure, let us introduce the quantity 
\begin{equation}
\label{eq:new}
\Sigma_{E'', E'}(\{\vc r_i\}) = \frac {(E' - U(\{\vc r_i\}))^{n / 2 - 1}} {(E''
- U(\{\vc r_i\}))^{n / 2 - 1}} \Theta(E' - U(\{\vc r_i\}))
\end{equation}
with $E' \leq E''$. We may then write
\begin{equation}
\label{eq:delta2}
\Delta_{E'}^{E''} S = -k \ln \langle \Sigma_{E'', E'}(\{\vc r_i\}) \rangle_{E''}
\end{equation}
The proof of this equation follows directly from the substitution of
eq.~(\ref{eq:new}) into eq.~(\ref{eq:delta2}), whence eq.~(\ref{eq:norm}) can
be identified and, after using the inverse proportionality between $C_{E}'$ and
$\omega(E)$, this leads to eq.~(\ref{eq:planck}) in difference form. Clearly,
both eqs~(\ref{eq:delta1}) and (\ref{eq:delta2}) may be used to calculate the
entropy difference. Shortly, we will consider the question of which function is
the most efficient from a computational perspective.

The above equations are to be used when the total energy is the only integral
of motion in the mechanical system. For completeness, we note the form that the
corresponding sigma functions must take when the averaging is done by molecular
dynamics means, if the objective is to obtain the density of states. In this
case, the $\sigma$ function becomes
\begin{equation}
\widetilde \sigma_{E, E_\mathrm{r}}(\{\vc r_i\}) = \frac {\Theta(E_\mathrm{r} -
U(\{\vc r_i\}) - \frac {\vc P^2} {2 M} )} {(E - U(\{\vc r_i\}) - \frac {\vc
P^2} {2 M} )^{n / 2 - 1}}.
\end{equation}
and the $\Sigma$ function is to be replaced by,
\begin{eqnarray}
\widetilde \Sigma_{E'', E'}(\{\vc r_i\}) & = & \frac {(E' - U(\{\vc r_i\}) -
\frac {\vc P^2} {2 M})^{n / 2 - 1}} {(E'' - U(\{\vc r_i\}) - \frac {\vc P^2} {2
M})^{n / 2 - 1}} \nonumber \\
& \times & \Theta \left (E' - U(\{\vc r_i\}) - \frac {\vc P^2} {2 M} \right )
\end{eqnarray}
Once again, heed must be paid to the value of $n$ so that the subtraction of
the center-of-mass momentum degrees of freedom is accounted for. In every other
respect, the equations for the entropy differences remain formally unchanged.
Quite conceivably, one might want to extract the corresponding entropy of the
system without these additional first integrals, in which case one may
introduce the function,
\begin{eqnarray}
\widetilde \Sigma'_{E'',E'}(\{\vc r_i\}) & = & \frac {(E' - U(\{\vc
r_i\}))^{n'/2 - 1}} {(E'' - U(\{\vc r_i\}) - \frac {\vc P^2} {2 M} )^{n / 2 -
1}} \nonumber \\
& \times & \Theta \left (E' - U(\{\vc r_i\}) \right )
\end{eqnarray}
and calculate entropies by eq.~(\ref{eq:delta2}). Here $n'$ exceeds $n$ by the
number of Cartesian components in the center-of-mass momentum. The
corresponding form for eq.~(\ref{eq:delta1}) follows by analogy.

We now turn to an analysis of the relative computational merits of
eqs~(\ref{eq:delta1}) and (\ref{eq:delta2}). To obtain $S(E)$ as a
(quasi-)continuous function of $E$, the calculation may
be subdivided into $N$ discrete energy segments over a predefined energy range.
For instance, if the energy interval $[E', E'']$, is subdivided into $N$
segments separated at energies $\{E_i\}$ such that $E_0 = E', E_i < E_{i+1}$;
$i=0, 1, \ldots, N-1$; and $E_N = E''$, then the total entropy difference is
given as a sum over the individual entropy differences for each segment,
\begin{equation}
\label{eq:segment}
\Delta_{E'}^{E''} S = \sum_{i=0}^{N - 1} \Delta_{E_i}^{E_{i+1}} S
\end{equation}
With eq.~(\ref{eq:delta1}), $N + 1$, and with eq.~(\ref{eq:delta2}), $N$
averages are needed. This difference becomes negligible for large $N$, which 
often corresponds to the most interesting situations. In the limit $N \to
\infty$, keeping $E'$ and $E''$ fixed, this gives $S(E)$ as a continuous
function of $E$ in the interval $[E', E'']$. We now note that as $N \to
\infty$, $E_{i+1} - E_i \to 0$ and in this limit, $\Sigma_{E_{i+1},E_i}(\{\vc
r_j\}) \to 1$ for all $\{\vc r_j\}$ accessible in the microcanonical ensemble
(that is, for all $\{\vc r_j\}$ such that $U(\{\vc r_j\}) \leq E_{i+1}$) and
because the $\Sigma$ function being averaged becomes identically unity, the
average $\langle \Sigma_{E_{i+1},E_i}(\{\vc r_j\})\rangle_{E_{i+1}}$ loses all
statistical uncertainty. Before we continue, we note that this quality is not
assured for the averages over the corresponding $\sigma$ functions, as they do
not enjoy the same guarantee, and even less so their ratio. 

In order to complete and strengthen the general argument, we should sum up and
consider the uncertainties of all the $N$ individual averages. Therefore,
considering the rate by which the averages of the $\Sigma$ functions approach
unity (and lose their statistical uncertainty), is of importance. Rearranging
eq.~(\ref{eq:segment}), it is clear that
\begin{equation}
\label{eq:converge}
\ln \langle \Sigma_{E_{i+1},E_{i}} \rangle_{E_{i+1}} \sim \frac
{\Delta_{E'}^{E''} S} {k N}
\end{equation}
In other words, the logarithm of each individual $\Sigma$ average tends to zero
inversely proportionally to $N$. In the numerical implementation, errors
will accrue if the ratio on the right-hand side becomes of the order of the
numerical precision. To keep the notation as simple as possible, we temporarily
restrict our attention to $n = 2$, in which case the formulae are
drastically simplified. In this case, for instance, the uncertainty of the
averages may be estimated from $\langle \Theta(E_i - U(\{\vc r_j\}))
\rangle_{E_{i+1}}$. For $N$ large enough, the converged value of $\langle
\Theta(E_i - U(\{\vc r_j\})) \rangle_{E_{i+1}}$ will be very close to, but
slightly less than, unity. With finite statistics, we estimate this value to be
$\alpha_i$. Because of our choice of $n =
2$, the average in question is composed only of terms being either unity or
zero. If there are $M_i$ terms equal to unity and $m_i$ terms equal to zero
sampled in the numerical averaging, then $\alpha_i = M_i / (M_i + m_i)$. In the
``worst case scenario'', the statistics of the ensemble averaging is so poor
(because of $M_i + m_i$ being chosen too small) that $\alpha_i$ is virtually a
non-uniform random number between zero and one. This value is thus different
from the actual converged value, which we denote $\beta_i$. The average
magnitude of this error is a measure of the uncertainty in the averaging. The
variance of the relative error, for instance, can be formulated as
\footnote{For the sake of notational simplicity, we consider only the Monte
Carlo probability distribution here. The argument is completely analogous in
the molecular dynamics case.},
\begin{eqnarray}
\mathrm{Var} \left (\frac {\alpha_i} {\beta_i} \right ) \equiv
\left \langle \frac {\alpha_i^2} {\beta_i^2} \right \rangle_{E_{i+1}} - 1 & = & \nonumber \\
\int \mathrm d \{\vc r_j\} \widehat W_{E_{i+1}}(\{\vc r_j\}) \frac {\Theta(E_i
- U(\{\vc r_j\}))} {\beta_i^2} & - & 1
\label{eq:variance}
\end{eqnarray}
Inserting the expression for $\widehat W_{E_{i+1}}$, we find that the integral
on the right-hand side is,
\begin{eqnarray}
\int \mathrm d \{\vc r_j\} \widehat W_{E_{i+1}}(\{\vc r_j\}) \Theta(E_i
- U(\{\vc r_j\})) & = & \nonumber \\
\frac {\int \mathrm d \{\vc r_j\} \Theta(E_i - U(\{\vc r_j\}))} {\int \mathrm d
\{\vc r_j'\} \Theta(E_i + \Delta_N E - U(\{\vc r_j'\}))} & & 
\label{eq:disc}
\end{eqnarray}
where we have introduced $\Delta_N E = (E'' - E') / N$.

We cannot hope to solve the integral in eq.~(\ref{eq:disc}) in the general
case, and like this obtain the variance as an explicit function of $N$. There
are, however, some conclusions to be drawn from the general form of the
right-hand side. In molecular systems, the accessible configuration space
generally increases superlinearly with increasing potential energy. Hence, the
integral in the denominator of eq.~(\ref{eq:disc}) should increase
superlinearly with increasing $\Delta_N E$. It follows immediately, that the
variance according to eq.~(\ref{eq:variance}) should decrease superlinearly
with decreasing $\Delta_N E \propto N^{-1}$ or, in other words, 
\begin{equation}
\frac {\int \mathrm d \{\vc r_j\} \Theta(E_i - U(\{\vc
r_j\}))} {\int \mathrm d \{\vc r_j'\} \Theta(E_i + \Delta_N E - U(\{\vc
r_j'\}))} = 1 + \mathcal O(N^{-a})
\end{equation}
where $a > 1$ is undetermined (but assuredly greater than unity). Thus, the
total variance (given as $N$ times the individual variance) will decrease to
zero as $\mathcal O(N^{1-a})$ when $N \to \infty$.  It follows that for a
sufficiently finely meshed energy grid, the $\Sigma$ function will always be
computationally more efficient than the $\sigma$ function, regardless of the
complexity of the system, \emph{as long as its accessible configuration space
increases superlinearly with increasing potential energy}. The general
argument, but with clumsier notation, can be carried through also with $n \neq
2$.  

In the numerical implementation, the limit $N \to \infty$ may of course not be
reached exactly and so a superior computational efficacy in the numerically
very demanding $N \to \infty$ limit is not necessarily relevant in actual
calculations. We must therefore also consider the relative efficacy of
eqs~(\ref{eq:delta1}) and (\ref{eq:delta2}) for finite energy differences.

As discussed by Davis \cite{davis11}, in the case of eq.~(\ref{eq:delta1}), the
constant $E_\mathrm{r}$ must be chosen so that both averages under the
logarithm are calculated with enough statistics. Too small values of
$E_\mathrm{r}$ restrict the statistics sampled, as not enough sampled
configurations do then have potential energies $U(\{\vc r_i\}) \leq
E_\mathrm{r}$. At the same time $E_\mathrm{r}$ must be less than or equal to
the smallest of the two energies for which the entropy difference is
calculated, meaning that a too large energy gap will be detrimental to the
statistics of the higher energy average. This essentially introduces an upper
bound for the energy difference for which the entropy difference can be
reliably calculated. As noted by Davis, this upper bound will depend on the
size of the system, because the fluctuations in potential energy become
smaller, the larger the system is (in the sense of the value of $n$). We note
that a similar restriction (for the same reasons) applies to
eq.~(\ref{eq:delta2}), in which $E'$ takes the place of $E_\mathrm{r}$. 

\begin{figure}
\begin{center}
\includegraphics{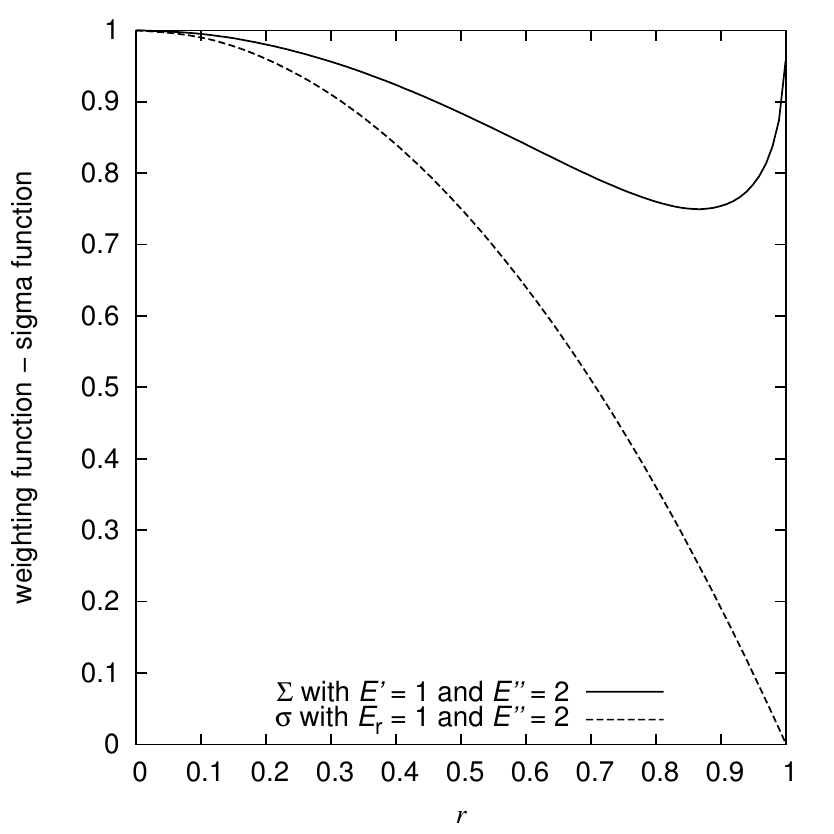}
\end{center}
\caption{Plot of the difference between either of the two sigma functions
for specific and arbitrarily chosen $E'' = 2, E' = E_\mathrm r = 1$ and $U(r) =
r^2$ and the importance sampling function (with $n = 3$) used in microcanonical
Monte Carlo simulations. In this plot, $C_{E''}'$ is arbitrarily set to
$C_{E''}' = 1$.} 
\label{fig:grad}
\end{figure}

To consider the question of convergence of the averages in more detail we
restrict our attention somewhat and assume that the microcanonical statistics
are sampled by importance sampling Metropolis Monte Carlo according to the
probability function $\widehat W_E(\{\vc r_i\})$ \footnote{This analysis does
hence not apply to the question of which sigma function is the most efficient
in molecular dynamics}.  In this case, a statistically good estimate of the
average of a function $A(\{\vc r_i\})$ is obtained if $A(\{\vc r_i\})$
contributes appreciably in regions where $\widehat W_E(\{\vc r_i\})$ is large,
and likewise contributes negligibly in regions where $\widehat W_E(\{\vc
r_i\})$ is close to zero. The question thus reduces to which of the two sigma
functions is most ``similar'' to $\widehat W_E(\{\vc r_i\})$, in the sense that
they share the domains where they are both of appreciable magnitude. For
instance, consider the ratios between the sigma functions and the Markov
weighting function,
\begin{eqnarray}
\label{eq:sim0}
\frac {\Sigma_{E'',E'}(\{\vc r_i\})} {\widehat W_{E''}(\{\vc r_i\})} & = & \frac {\left (E' - U(\{\vc r_i\})
\right )^{n/2-1} } {C_{E''}' \left (E'' - U(\{\vc r_i\})
\right )^{n - 2}} \nonumber \\
& & \frac {\Theta \left(E' - U(\{\vc r_i\}) \right )} {\Theta \left (E'' -
U(\{\vc r_i\}) \right)}, 
\end{eqnarray}
\begin{equation}
\label{eq:sim1}
\frac {\sigma_{E'',E_\mathrm{r}}(\{\vc r_i\})} {\widehat W_{E''}(\{\vc r_i\})}
= \frac {\Theta(E_\mathrm{r} - U(\{\vc r_i\}))} {C_{E''}' (E'' - U(\{\vc
r_i\}))^{n - 2}},
\end{equation}
\begin{equation}
\label{eq:sim2}
\frac {\sigma_{E',E_\mathrm{r}}(\{\vc r_i\})} {\widehat W_{E'}(\{\vc r_i\})} =
\frac {\Theta(E_\mathrm{r} - U(\{\vc r_i\}))} {C_{E'}' (E' - U(\{\vc r_i\}))^{n
- 2}}.
\end{equation}
The less similar the two functions are, the less constant is their ratio. In
the simplest case, the limiting case of an ideal gas, the $\{\vc r_i\}$
gradients vanish for all of these ratios and the relative qualities of the
importance sampling of the averages are not distinguishable between the
$\sigma$ and $\Sigma$ functions. When interactions are present, this is no
longer the case. Whereas the resulting ratios of both eqs.~(\ref{eq:sim1})
and~(\ref{eq:sim2}) consist of a practically constant numerator and a
monotonously and smoothly decreasing denominator as a function of $U(\{\vc
r_j\})$, the ratio between the $\Sigma_{E'',E'}$ function and $\widehat
W_{E''}$ presents a smoothly decreasing function of $U$ for both numerator and
denominator. Hence, it would seem that this ratio is more invariant with respect
to changes in $\{\vc r_j\}$ (and hence $U(\{\vc r_j\})$) than the others. This
becomes, once again, particularly pronounced when $E' \to E''$. A more-or-less
constant difference can be used just as well as an indication of similarity
and this is what we consider in Fig.~\ref{fig:grad} in the case of a
three-dimensional harmonic oscillator for which $U(r) = r^2$. This is a model
potential for atomic crystals and makes for a reasonably relevant comparison.
As anticipated, the difference $\Sigma_{E'', E'} - W_{E''}$ exhibits much less
variation than $\sigma_{E'',E'} - W_{E''}$. Above $E = E' = E_\mathrm r$, they
become identical.

We do not offer any numerical experiments to illustrate the method. This has
already been achieved by Davis \cite{davis11} on a non-trivial system using the
$\sigma$ function. However, the numerical upper limitation on the size of
tractable systems that Davis points out is nonetheless important to recall. A
similar limitation, although much less severe, is present already in the
microcanonical sampling algorithm \cite{severin78, schranz91, ray91}, as
acceptance probabilities for a trial move taking the system from potential
energy $U'$ to $U''$ at total energy $E$ are proportional to the ratio
\begin{displaymath}
\left (\frac {E - U''} {E - U'} \right )^{n/2-1} \Theta(E - U'')
\end{displaymath}
which for large $n$ values ought to become difficult for the computer
architecture to resolve to sufficient accuracy, as the acceptance ratio takes
on a more ``step function''-like form. Severin \textit{et al.} \cite{severin78}
initially introduced the sampling algorithm for sampling the internal degrees
of freedom of single molecules. Obtaining the density of states of complicated
polyatomics, needed for instance in statistical reaction rate theories, is thus
a natural application of a method such as this. Nevertheless, Ray \cite{ray91}
reports comfortable simulations on up to 500 particles, using this
microcanonical sampling. Such system sizes should be sufficient for many
purposes in statistical mechanics.

In conclusion, we note one interesting formal property of the $\Sigma$
averages: from a single microcanonical molecular dynamics (or Monte Carlo) run,
\emph{in principle} the entire $S(E)$ function is obtainable (up to an additive
constant). This follows since the energy $E'$ is arbitrary in
eq.~(\ref{eq:delta2}), yet does not affect the dynamics. Nevertheless, it is
clear from the limitations discussed above, that good statistics would only be
achieved in a narrow range below $E''$. However, for very small systems, this
range might be quite broad.


%

\end{document}